\documentclass[journal,onecolumn,11pt]{IEEEtran}

\usepackage{ifpdf}
\ifpdf
	\usepackage[pdftex]{graphicx}
	\graphicspath{{figure-pdf/}}
	\DeclareGraphicsExtensions{.pdf, .jpg, .png, .tif}
\else
	\usepackage[dvips]{graphicx}
	\graphicspath{{figure-eps/}}
	\DeclareGraphicsExtensions{.eps, .ps}
\fi

\usepackage[cmex10]{amsmath}
\usepackage{amssymb}
\usepackage{url}

\usepackage[caption=false, font=footnotesize]{subfig}
\usepackage{algorithm, balance}
\usepackage{multirow}

\newtheorem{theorem}{Theorem}[section]
\newtheorem{proposition}[theorem]{Proposition}

\begin{document}
%
\title{Multiple Configurations LT Codes}
%
%
%

\author{Pei-Chuan~Tsai,
        Chih-Ming~Chen,
        and~Ying-ping~Chen,~\IEEEmembership{Member,~IEEE}
\thanks{The authors are with the Department of Computer Science, National Chiao Tung University, Hsinchu, TAIWAN.}
\thanks{Email addresses: \{pctsai, ccming, ypchen\}@nclab.tw.}}

%


\maketitle

\begin{abstract}
This paper introduces a new scheme of LT codes, named \textit{multiple configurations}. In multiple configurations LT codes (MC-LT codes), multiple sets of output symbols are simultaneously provided to receivers for recovering the source data. Each receiver, without the need to send information back to the sender, is capable of receiving the output symbols generated by some configuration chosen according to its own decoding phase. Aiming at the broadcasting scenarios without feedback channels, the proposed MC-LT codes are shown to outperform the optimal pure LT codes at the cost of encoding and transmitting units. In this paper, the inspiration of MC-LT codes is presented, how MC-LT codes work is described by giving examples, in which the optimal pure LT codes are outperformed, and a practical design of MC-LT codes, which is analytically proved to have at least the same performance bound as the pure LT codes, is proposed. The results of numerical simulation experiments demonstrate that the proposed practical design of MC-LT codes can deliver better performance than the LT codes in comparison. In summary, this paper creates new potential research directions for LT codes, and MC-LT codes are a promising variant of LT codes, especially for broadcasting scenarios.\par
\end{abstract}


%
\IEEEpeerreviewmaketitle

\section{Introduction}
\label{sec:Introduction}

Luby transform (LT) codes~\cite{Luby:02:inproceedings} proposed by Michael Luby are a well-known practical implementation of digital fountain codes~\cite{Byers:98:inproceedings}. One of the key characteristics of LT codes is \textit{ratelessness}, which allows a sender to generate unlimited output symbols without a fixed code rate. The receivers that are interested in obtaining the source data can continuously listen to the output symbols and reconstruct source data once a sufficient amount of output symbols has been received. Ratelessness is an advantage of LT codes, with which a sender does not need to attend to the individual differences of receivers, and the performance of LT codes is independent of channel erasure rate because of another key characteristic, the \textit{universal property}. LT codes, therefore, can be adopted as a solution for transmission through an unknown channel or in broadcasting scenarios over a heterogeneous network~\cite{He:10:inproceedings,Ji:11:inproceedings}.\par

Because the transmission overhead of LT codes highly depends on the degree distribution applied to generate output symbols, the proposal of LT codes introduced a practical degree distribution for an arbitrary code length. The distribution was named \textit{robust soliton distribution}, of which the performance had been confirmed by theoretical analysis: When $k + {\cal O}(\ln^{2} (k / \delta) \sqrt{k})$ output symbols are received, the source data can be recovered with a probability at least $1-\delta$, where $k$ is the code length and $\delta$ is a parameter of the adopted robust soliton distribution. Robust soliton distribution was designed from the viewpoint of the sender side to provide a general set of output symbols which assures receivers to be able to recover the data with a promising probability regardless of the channel characteristics of each receiver. However, the authors of this paper observed that the utility of output symbols of different degrees actually varies along with the progress of decoding. As a receiver, collecting more output symbols of higher utility as decoding progresses would help the recovery of source data. Hence, the performance of LT codes may be improved by dynamically adjusting the degree proportion of output symbols.\par

To the best of the authors' limited knowledge, \cite{Bonello:09:article} was the only published work based on this idea, called \textit{reconfigurable rateless codes}. The approach proposed in~\cite{Bonello:09:article} requires a feedback channel between the sender and the receiver for assessing the current channel state and accordingly adapts the LT code degree distribution at the sender side to facilitate the transmission to the receiver. The setting of communicating via a feedback channel and adjusting output symbols for one or a few receivers renders the proposed approach inappropriate for the scenarios of broadcasting or with a lot of receivers on channels with a wide-ranged erasure rates.\par

In this paper, mainly aiming to tackle the broadcasting scenarios in which no feedback channel exists between senders and receivers, we propose a new scheme of LT codes called \textit{multiple configurations}. In multiple configurations LT codes (MC-LT codes), the performance is improved by simultaneously providing multiple sets of potential output symbols. Receivers can individually, independently choose the most helpful set of output symbols according to their own local decoding status. The cost of the obtained performance is the number of encoding and transmission units, which is usually negligible for broadcasting scenarios. The fundamental of this work is based on the observation that the degree of the most helpful output symbols would vary at different decoding status, but the sender cannot adapt the degree distribution for some small portion of receivers. If the receivers are able to collect more helpful output symbols corresponding to their own decoding status, better performance of recovering source data can be achieved.\par

In this paper, MC-LT codes will be proposed and shown to outperform the optimal pure LT codes. We will present the inspiration of MC-LT codes, describe how MC-LT codes work by giving examples in which the optimal pure LT codes are outperformed, and propose a practical design of MC-LT codes, which is theoretically proved to have at least the same performance bound as the pure LT codes with robust soliton distribution. Our simulation experimental results demonstrate that the proposed design of MC-LT codes can deliver better performance than the LT codes in comparison for a real-time multimedia streaming scenario.\par

The remainder of the paper is organized as follows. To learn the needs of desired degrees of output symbols at each decoding phase, we give the definition and the detailed analysis of utility degree of an output symbol in section~\ref{sec:DegreeUtility} with the derivation of the domination relation between output symbols of different degrees. Based on the analysis, we propose the multiple configurations LT codes to enable receivers to choose a desirable set of output symbols in section~\ref{sec:New_schemes}. The improvement and performance bound are also shown in section~\ref{sec:New_schemes}. The experimental results and the comparison with the original LT codes are presented in section~\ref{sec:Experiments}. Finally, section~\ref{sec:Conclusions} concludes this paper by presenting the contributions of this work and certain potential research direction for future work.\par


\section{Study of utility degrees}
\label{sec:DegreeUtility}

\begin{figure}[t]
\centerline{\includegraphics[width = 90mm]{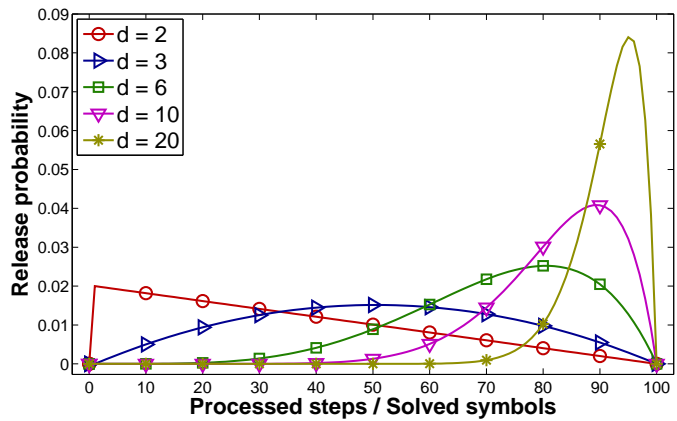}}
\caption{The figure plots the release probabilities of output symbols of different degrees $d$ under different decoding phases and the code length $k$ is $100$.}
\label{fig:DegreeRelease}
\end{figure}

In this section, we begin with the observation of the release probability of output symbols. Symbol release~\cite{Luby:02:inproceedings} was firstly introduced by Luby to illustrate that the degree of an output symbol would be reduced to one and the released symbol could be used to recover some other unsolved input symbols. The probability of symbol release, denoted as release probability, is a computable function with two arguments, degree of an output symbol and the number of processed input symbols. Figure~\ref{fig:DegreeRelease} displays a plot of release probability for output symbols of different degrees. The plot shows that output symbols of lower degrees will be released in early decoding phases and those of higher degrees will be released later. Moreover, the curves roughly reveal the contribution of different degrees at different stages for a successful decoding. In Figure~\ref{fig:DegreeRelease}, it is clear that an output symbol of degree $20$ is more useful than the others when $90$ input symbols have been solved. The observable fact gives the inspiration of this work that it must be advantageous if receivers are enabled to collect output symbols of preferred degrees in different decoding phases. Next, we make a further investigation on the preference between different output symbol degrees.\par

\subsection{Utility degrees}
\label{subsec:utilitydegree}

For expressing the contribution of an output symbol, we examined the decoding process and found that as the decoding progress continues, more and more input symbols are solved. A received output symbol may be useless when all the input symbols composing it, which are called \textit{neighbors} in LT codes, have been recovered. Considering when an output symbol of degree $d$ reaches some receiver, there are $i$ neighbors of the output symbol still covered and able to be used for further decoding. We denote the number $i$ as \textit{utility degree} to represent the useful information remaining in this output symbol and use $D(d, i, u)$ to represent the probability of the output symbol of degree $d$ containing $i$ utility degrees when there are $u$ unsolved input symbols. Therefore, the probability $D(d, i, u)$ follows a hypergeometric distribution as
\begin{equation}
    D(d,i,u) = \frac{ {u\choose i}\cdot{{k-u}\choose{d-i}} }{{k\choose d}} \;, \\
    \mbox{where} \; \max(0 , d-k+u) \leq i \leq \min(d , u) \; . \nonumber
    \label{eq:hypergeomtric}
\end{equation}
An output symbol with any nonzero utility degree is useful because it must be released at some time after the current decoding step. On the contrary, an output symbol is useless if the utility degree decays to zero. Receiving such output symbols causes redundancy in data transmission.\par

\subsection{Decay of utility degrees}
\label{subsec:DegreeDecay}

\begin{figure}[t]
\centerline{\includegraphics[width = 90mm]{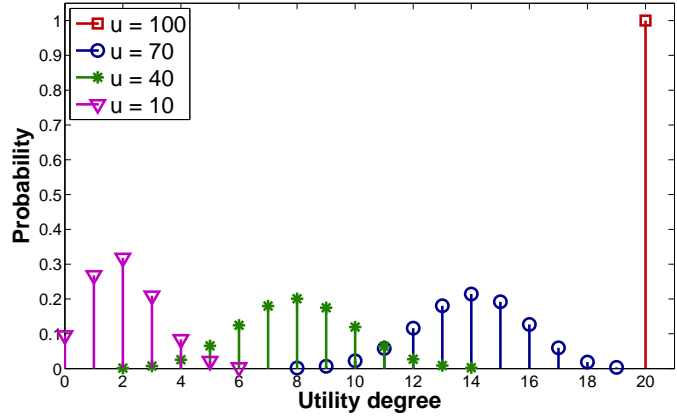}}
\caption{Assume that $k=100$, and an output symbol of degree $20$ is received. The figure illustrates the utility degree distributions at four different moments, denoted by different $u$ which represents the number of unsolved input symbols, in the decoding process.}
\label{fig:DegreeDecay}
\end{figure}

\begin{figure}[t]
\centerline{\includegraphics[width = 90mm]{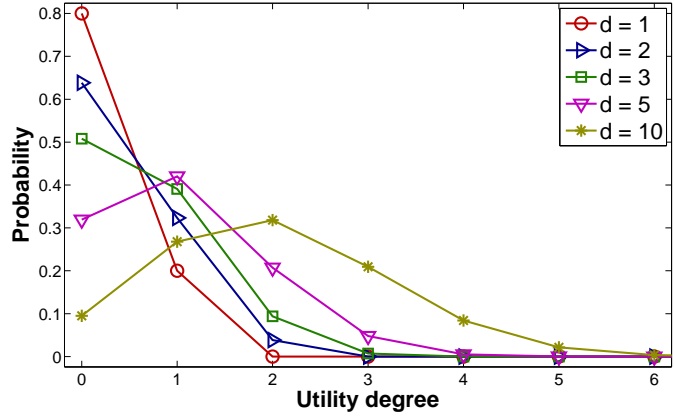}}
\caption{Assume that $k=100$, and $80$ input symbols have been solved. The figure illustrates the utility degree distributions of received output symbols of different degrees. The curves of $d=1$, $d=2$, and $d=3$ are dominated by $d=5$ except for the probability on the useless zero degree.}
\label{fig:DegreeDomination}
\end{figure}

It can be imagined that the utility degree of an output symbol would slowly decay as the decoding process continues. Figure~\ref{fig:DegreeDecay} displays several instances of the utility degree distribution of an output symbol of degree $20$ for $k=100$. The figure clearly illustrates that the utility degree decreases while more and more input symbols are solved. Under the circumstances, a feasible solution to reduce the transmission redundancy is to change the degree proportion of received symbols since symbols of higher degrees are preferred in the situation. Another illustration is presented in Figure~\ref{fig:DegreeDomination}, in which $k=100$, and $80$ input symbols have been solved. The utility degree distributions of output symbols of different degree $d$ are illustrated. It can be found that the output symbols of degrees $d=1$, $d=2$, and $d=3$ would become useless, i.e., with zero utility degree, with high probability. Moreover, the probabilities of nonzero utility degree in distributions, $d=1$, $d=2$, and $d=3$, are all lower than that in distribution with degree $d=5$. These degrees $d=1$, $d=2$, and $d=3$ are dominated by the degree $d=5$ in such a situation. Hence, it is definitely advantageous in the case to replace output symbols of degree $d\le3$ by $d=5$. We must note that distributions of lower degrees are not always dominated by greater degrees such as the relation between $d=2$ and $d=10$ in the Figure~\ref{fig:DegreeDomination}.\par

\subsection{Domination of utility degrees}
\label{subsec:DegreeDomination}

According to the observation in the previous section, a substitution is definitely advantageous if there is a dominative relationship between two degrees. The dominative relationship can be formulated as
\begin{equation}
    D(d_2,i,u) \geq D(d_1,i,u) \text{  for  } i>0 \; ,
\label{eq:dom}
\end{equation}
in which there are $u$ unsolved input symbols and the probability of $d_2$ is always larger than $d_1$ for any nonzero utility degree $i$. Let $d_1=d$ and $d_2=d+1$, we obtain the condition of domination in the particular case by reducing the Equation~(\ref{eq:dom}) as
\begin{equation}
    \frac{ {u\choose i}\cdot{{k-u}\choose{d+1-i}} }{{k\choose {d+1}}} \geq \frac{ {u\choose i}\cdot{{k-u}\choose{d-i}} }{{k\choose d}}
    \nonumber\\
    \frac{(k-d-1)!(d+1)!}{(k-u-d-1+i)!(d+1-i)!} \geq \frac{(k-d)!d!}{(k-u-d+i)!(d-i)!}
    \nonumber\\
    (d+1)(k-u-d+i) \geq (k-d)(d-i+1)
    \nonumber\\
    u \leq \frac{i\cdot(k+1)}{d+1} \text{  for  } i>0 \; .
\label{eq:DegreeDominationRelation}
\end{equation}

From the derivation result, choosing $u=(k+1)/(d+1)$ obviously satisfies the Equation~(\ref{eq:DegreeDominationRelation}) for any integer $i>0$. Thus, the utility degree distribution of degree $d$ is dominated by degree $d+1$ when the number of unsolved input symbols is less than $(k+1)/(d+1)$, i.e., after the decoding step $k-(k+1)/(d+1)$.\par

\section{New schemes of LT codes}
\label{sec:New_schemes}

\begin{figure}[t]
\centerline{\includegraphics[width = 90mm]{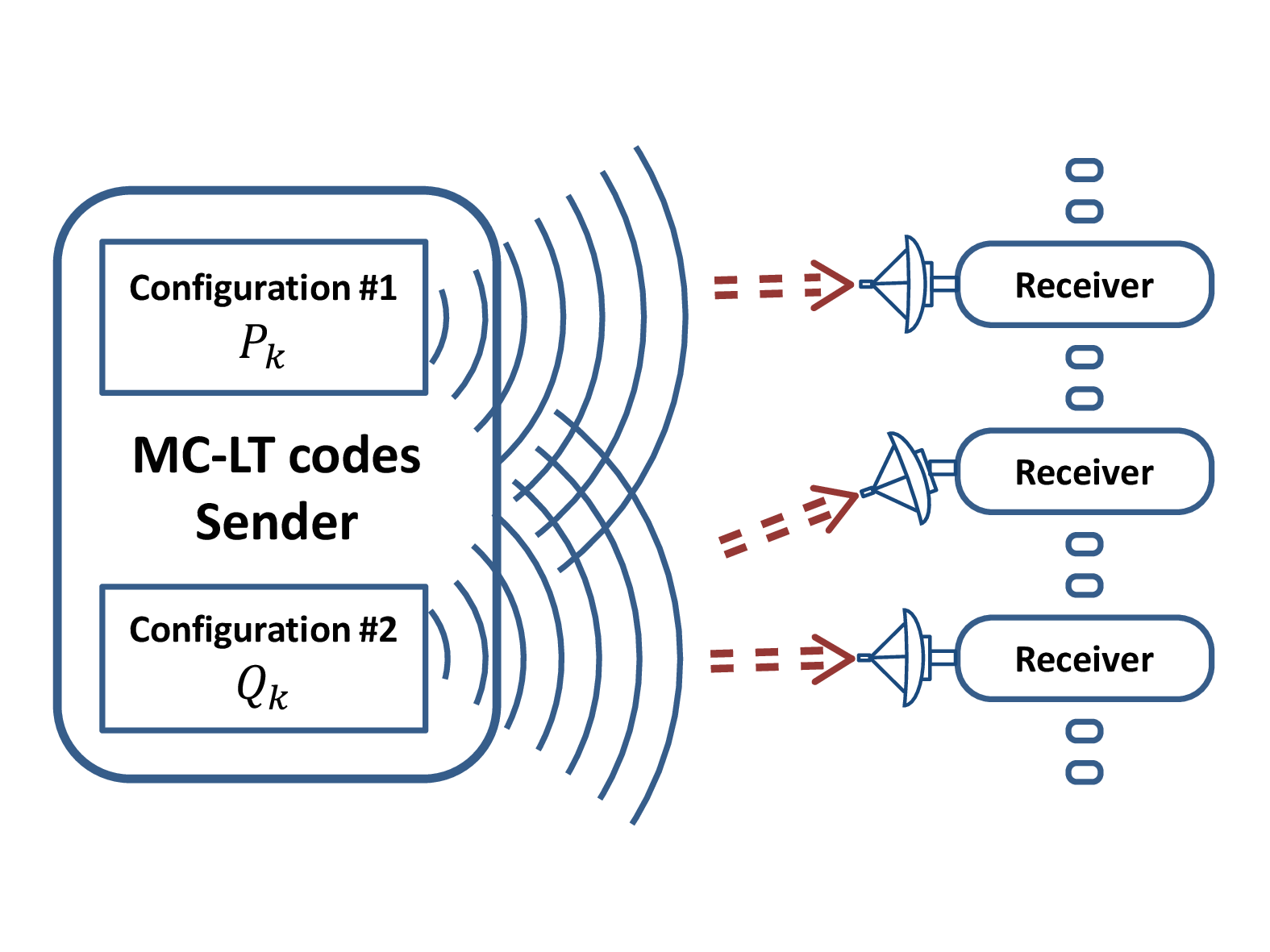}}
\caption{The figure illustrates a 2-configuration LT codes. The two configurations are built by referring the different degree distributions, $P_k$ and $Q_k$. Two configurations serve receivers at the same time. Receivers can individually, independently switch the data source to receive desired output symbols, and no feedback channel to the sender is required.}
\label{fig:MC-LT}
\end{figure}

Based on the observation and derivation presented in section~\ref{sec:DegreeUtility}, we propose a new scheme of LT codes, called \textit{multiple configurations LT codes} (MC-LT codes). The goal of this scheme is to make receivers capable of collecting more helpful output symbols without any feedback channels. The receivers able to collect more useful output symbols for recovering source data will definitely get better performance. Using feedback channels for getting preferences of receivers highly restricts the application of LT codes in two ways: 1) Feedback channels must exist. 2) In broadcasting scenarios, it is highly unlikely for sender to satisfy all receivers by adjusting the degree distribution. Therefore, we design the scheme for improving the data transmission performance with keeping all the advantages of original LT codes. In our scheme, the sender simultaneously generates multiple sets of output symbols according to multiple configurations. Each receiver can choose which set of output symbols to receive depending on its own decoding stage. Thus, no feedback channel between sender and receiver is required, and the advantages of LT codes such as the universal property and being suitable for broadcasting scenarios are retained. Figure~\ref{fig:MC-LT} illustrates an example of MC-LT codes with two configurations. In the example, MC-LT codes generate two sets of output symbols and transmit them simultaneously. The receivers in the scenario can individually, independently receive any one of the two sets of output symbols according to the local decoding stages. The configuration represents a set of parameters that can control the encoding procedure for generating output symbols, which can be the adopted degree distribution, the sampling mechanism for degree $d$ or other feasible variants of LT codes. In this work, we will focus on designing the degree distributions adopted in 2-configuration LT codes for simplicity and illustration, although the concept and architecture of MC-LT codes are not limited to only 2 configurations. In this section, we will use a basic example to demonstrate the performance improvement achievable by using MC-LT codes over the pure LT codes and give a guideline for designing MC-LT codes.\par

\begin{figure}[t]
\centering
\subfloat[LT codes]
	{\includegraphics[width=48mm]{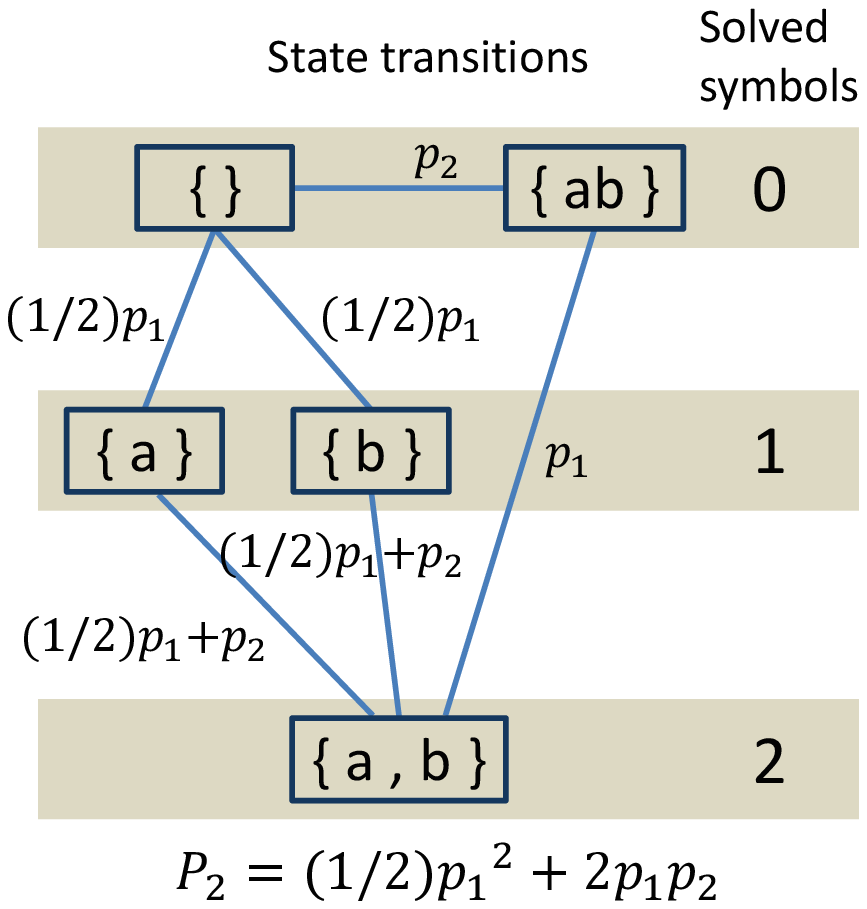}\label{subfig:pureLTcodes}}
\subfloat[MC-LT codes]
	{\includegraphics[width=48mm]{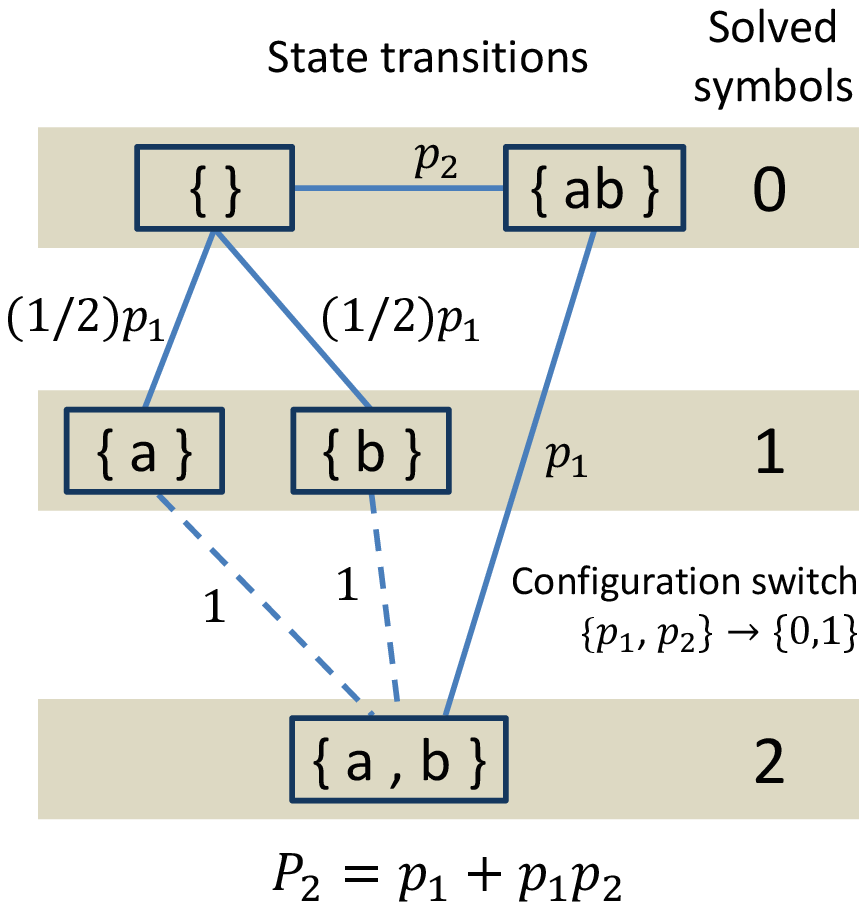}\label{subfig:MCLTcodes}}
\caption{The decoding state transitions are completely displayed for code length $k=2$. Suppose that the two input symbols are $a$ and $b$. The adopted degree distribution is $P_2=\{p_1,p_2\}$. The symbol sets in square nodes denote the current information about input symbols and the shaded rows mark the number of solved input symbols, which is the critical factor of configuration switch.}
\label{fig:toyexample}
\end{figure}

\subsection{Demonstration of MC-LT codes}
\label{subsec:DemoMCLT}

Figure~\ref{fig:toyexample} illustrates a simple example for demonstrating how MC-LT codes work. Here we compare the pure LT codes with MC-LT codes for a code length $k=2$. Figure~\ref{subfig:pureLTcodes} shows the complete state transitions for a successful decoding of LT codes with $k=2$. The presentation of the state transition was from~\cite{Hyytia:07:inproceedings}, which considered all possible decoding combinations to find the optimum degree distributions for short message length LT codes. In this example, we use the code length $k=2$ and the input symbols are $a$ and $b$. The symbol sets in square nodes represent the current decoding status of a receiver. $\{b\}$, for example, indicates the state that the input symbol $b$ has been recovered. $\{ab\}$ means that the output symbol consists of both input symbols has been received but no input symbol is solved. Assume the adopted degree distribution is $P_2 = \{p_1, p_2\}$, then the probability of each state transition can be shown in Figure~\ref{subfig:pureLTcodes}. Let $\mathcal{P}_n$ denote the successful decoding probability when exactly $n$ output symbols are received. Since we have the complete state transitions and the probability of each state transition, the probability $\mathcal{P}_2$ can be expressed as $\mathcal{P}_2 = (1/2)p_1^2 + 2p_1p_2$. Combined with the constraint that $\sum^k_{d=1}{p_d}=1$, we can optimize the term and get the maximum value of $\mathcal{P}_2 = 0.666$ for $P_2 = \{0.666, 0.334\}$, which is the optimal LT code for $k=2$, proved in \cite{Hyytia:07:inproceedings}.\par

On the other hand, Figure~\ref{subfig:MCLTcodes} shows the complete state transitions of MC-LT codes. According to the utility degree domination relation given by Equation~(\ref{eq:DegreeDominationRelation}), when the number of unsolved input symbols is less than $(k+1)/(d+1)$, the degree $d$ is dominated by degree $d+1$. Hence, we set the degree distribution of the second configuration as $\{p_1', p_2'\} = \{0, 1\}$, and the switch point is after one input symbol solved. By applying this setting, the probability of each state transition of MC-LT codes is expressed in Figure~\ref{subfig:MCLTcodes}. The dash lines show the differences from the pure LT codes, which would forward states, $\{a\}$ and $\{b\}$, to the full recovery. According to the probabilities of complete state transitions, the successful decoding probability of MC-LT codes can also be derived as $\mathcal{P}_2 = p_1 + p_1p_2$. We have the maximum value of $\mathcal{P}_2 = 1$ for $P_2 = \{1, 0\}$. The example shows a fact that the proposed scheme of multiple configurations definitely increases the maximum successful decoding probability, even in the case when the code length is extremely short. Next, we extend the code length and show a possible design of MC-LT codes.\par

\subsection{MC-LT codes for small code length $k$}
\label{subsec:SmallSizeMCLT}

\begin{table}[t!]
\caption{The table shows the results of the optimal pure LT codes and optimized 2-configuration LT codes for code length $k=3$.}
\begin{center}
    \begin{tabular}{|c||c|c|c|c|}
    \hline
    \multirow{2}{*}{Degree} &
    \multicolumn{1}{c|}{pure LT codes} &
    \multicolumn{2}{c|}{MC-LT codes} \\
    \cline{2-4}
	& $P_3$ & $P_3$ & $Q_3$ \\\hline
    $1$ & 0.517 & 0.701 & 0   \\ \hline
    $2$ & 0.397 & 0.299 & 0   \\ \hline
    $3$ & 0.086 & 0     & 1   \\ \hline\hline
    $\mathcal{P}_3$ & \multicolumn{1}{c|}{0.452} & \multicolumn{2}{c|}{0.742} \\ \hline
    \end{tabular}
    \label{t:MCLT-k3_optimization}
\end{center}
\end{table}

\begin{table}[t!]
\caption{The table shows the results of the optimal pure LT codes and optimized 2-configuration LT codes for code length $k=4$.}
\begin{center}
    \begin{tabular}{|c||c|c|c|c|}
    \hline
    \multirow{2}{*}{Degree} &
    \multicolumn{1}{c|}{pure LT codes} &
    \multicolumn{2}{c|}{MC-LT codes} \\
    \cline{2-4}
	& $P_4$ & $P_4$ & $Q_4$ \\\hline
    $1$ & 0.429 & 0.527 & 0   \\ \hline
    $2$ & 0.430 & 0.473 & 0   \\ \hline
    $3$ & 0.100 & 0     & 0   \\ \hline
    $4$ & 0.041 & 0     & 1   \\ \hline\hline
    $\mathcal{P}_4$ & \multicolumn{1}{c|}{0.315} & \multicolumn{2}{c|}{0.554} \\ \hline
    \end{tabular}
    \label{t:MCLT-k4_optimization}
\end{center}
\end{table}

As the way used by \cite{Hyytia:07:inproceedings} for finding the optimal degree distribution for small code length $k$, we can apply the same idea to design MC-LT codes for small code length $k$. Assume we use the two degree distributions $P_k = \{p_1, \ldots, p_k\}$ and $Q_k = \{q_1, \ldots, q_k\}$ to design a 2-configuration LT codes for code length $k$. We can find out the complete state transitions, set the switch point, and then derive the expression of successful decoding probability $\mathcal{P}_k$ to get optimized degree distributions $P_k$ and $Q_k$. For example, we set switch point as after $\lceil k - (k+1)/2 \rceil$ input symbols are solved for replacing degree $1$, i.e., for satisfying the degree domination relation $u \leq (k+1)/(d+1)$. Thus, for $k=3$, the successful decoding probability $\mathcal{P}_3$ can be expressed as
\begin{align}
    \mathcal{P}_3 &= \frac{2}{9}p_{1}p_{2}( 5p_{2} + 12p_{3} + 2q_{1} + 4q_{2} + 6q_{3}) \nonumber\\
		  &+ \frac{2}{9}p_{1}^{2}( 2p_{2} + 6p_{3} + q_{1} + 2q_{2} + 3q_{3}) \; .
\label{eq:FC_k3}
\end{align}
We can maximize the decoding probability $\mathcal{P}_3$ and get the optimal degree distributions of $P_3$ and $Q_3$ as listed in Table~\ref{t:MCLT-k3_optimization}. Using the same approach for optimizing 2-configuration LT codes for code length $k=4$ and the results are listed in Table~\ref{t:MCLT-k4_optimization}. We also list the optimal degree distributions of pure LT codes obtained by \cite{Hyytia:07:inproceedings} in both Tables~\ref{t:MCLT-k3_optimization} and \ref{t:MCLT-k4_optimization}. Since Hyyti\"{a} et al. exhaustively considered all possible decoding state transitions, for pure LT codes, it is impossible to make any performance improvement by refining the degree distribution. However, the results shown in Table~\ref{t:MCLT-k3_optimization}, Table~\ref{t:MCLT-k4_optimization}, and the previous section reveal that the proposed scheme of multiple configurations can break the limitation of pure LT codes and deliver performance better than the theoretical optimum by trading in the cost of multiple configurations, which in broadcasting scenarios is affordable or even negligible.\par

\subsection{MC-LT codes for arbitrary code length $k$}
\label{subsec:ArbitraryMCLT}

Although the method described in section~\ref{subsec:SmallSizeMCLT} can be used to build 2-configuration LT codes, and performance better than that of the pure LT codes can be obtained, we still face the difficulty that, for arbitrary code length $k$, it is hard to exhaustively list all possible decoding state transitions and derive the expression of the successful decoding probability $\mathcal{P}_k$ to optimize the settings of degree distributions. Therefore, in this work, we also propose a practical way to design MC-LT codes for arbitrary code length $k$. The idea is inspired from the robust soliton distribution proposed by Luby.\\

\noindent \textit{Robust soliton distribution} $\mu(d)$:
\begin{equation}
    \rho(d) =
    \left\{
    \begin{array}{lll}
        \frac{1}{k} & \text{for} & d = 1
        \\
        \frac{1}{d(d-1)} & \text{for}& d = 2, 3, \ldots, k
    \end{array}
    \right.
	\; ,
    \label{eq:rho}
\end{equation}
\[
    R = c \ln(k/\delta)\sqrt{k} \; ,
\]
\begin{equation}
	\tau(d)=
	\left\{
		\begin{array}{lll}
        R / dk & \text{for} & d = 1,\ldots,(k/R)-1
		\\
        \frac{R \ln (R/\delta)}{k} & \text{for} & d = k/R
        \\
        0 & \text{for} & d = (k/R) + 1,\ldots,k
		\end{array}
	\right.
    \; ,
    \label{eq:tau}
\end{equation}
\begin{eqnarray}
        \beta &=& \sum^{k}_{d=1} \left( \rho(d)+\tau(d) \right) \label{eq:beta} \; ,\nonumber\\
        \mu(d)&=& \frac{\rho(d)+\tau(d)}{\beta}\text{  for  } d = 1, \ldots, k \; .
\end{eqnarray}

In the design of robust soliton distribution, the distribution $\rho(d)$ is responsible for the early phase of decoding and the distribution $\tau(d)$ is for the late decoding phase. According to the theoretical analysis of robust soliton distribution, the distribution $\tau(d)$, especially the spike $\tau(k/R)$, would be more helpful than $\rho(d)$ during the late decoding phase. Based on the principle of MC-LT codes: providing useful output symbols for receivers to choose at different decoding phases, we re-model the robust soliton distribution into a pair of new distributions. The first distribution of the pair is called the \textit{starting distribution} (or \textit{starter}), and the other is called the \textit{closing distribution} (or \textit{closer}), where the names come from the terminology for baseball pitchers.\\

\begin{figure}[t!]
\centerline{\includegraphics[width=90mm]{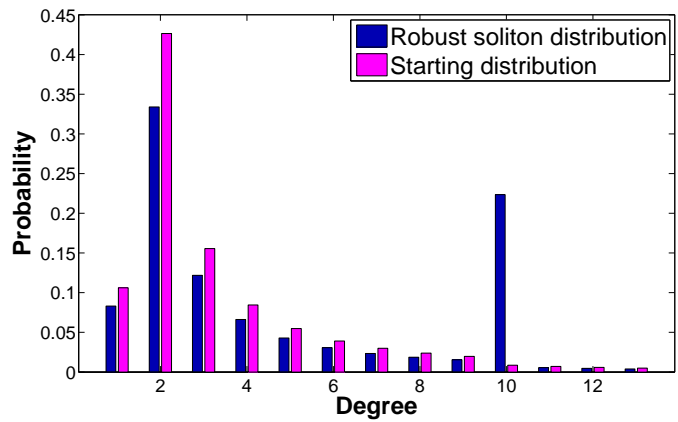}}
\caption{The figure shows a partial comparison between robust soliton distribution and starting distribution with code length $k=30$. The distribution parameters are $c=0.1$ and $\delta=0.1$.}
\label{fig:MC_dis}
\end{figure}

\textit{Starting distribution} $\lambda(d)$:
\begin{equation}
    \gamma = \sum^{k}_{d=1} \rho(d) + \sum^{(k/R)-1}_{d=1} \tau(d) \; ,\nonumber\\
    \lambda(d) =
    \left\{
    \begin{array}{lll}
      	\frac{\rho(d)+\tau(d)}{\gamma} & \text{for} & d = 1, \ldots, (k/R) - 1
        \\
        \frac{\rho(d)}{\gamma} & \text{for} & d = k/R, \ldots, k
    \end{array}
    \right.
	\; .
\end{equation}

\textit{Closing distribution} $\theta(d)$:
\begin{equation}
    \theta(d) =
        \left\{
        \begin{array}{lll}
            1 & \text{if} & d = k/R
            \\
            0 & \text{if} & d \neq k/R
        \end{array}
        \right.
    \; .
\end{equation}

The definitions of $\rho(d)$ and $\tau(d)$ are exactly the same as Equations (\ref{eq:rho}) and (\ref{eq:tau}). It can be clearly seen that each of the two distributions is a partial robust soliton distribution. The peak at the tail of robust soliton distribution is removed to form the closing distribution which always generates output symbols of the same degree $k/R$. The starting distribution is then exactly the remainder of robust soliton distribution after normalization. The proposed pair of distributions is used to construct the practical 2-configuration LT codes. Figure~\ref{fig:MC_dis} shows the difference between robust soliton distribution and starting distribution. After constructing the pair of degree distributions, the setting of switch point of the 2-configuration LT codes is also referred to the contributions of distributions $\rho(d)$ and $\tau(d)$ among the whole decoding process. The switch point is set as when $R$ input symbols remain unsolved since the distribution $\tau(d)$ was designed for covering the last $R$ input symbols. Therefore, in the proposed 2-configuration LT codes, the decoding process is divided into two phases and the starter is responsible for the first phase and the closer is for the other phase.\par

We then formally analyze the performance of the proposed 2-configuration LT codes. The following analysis is based on the framework presented in~\cite{Luby:02:inproceedings}. In the proof of performance of LT codes with robust soliton distribution, the decoding process can be divided into two phases: 1) decoding from the beginning to the status at which only $R$ input symbols remain unsolved; 2) decoding the last $R$ input symbols. For phase one, Luby showed the proposition that it has robust uniform release probability.\par

\begin{proposition}
  \emph{(robust uniform release probability)}
  \label{pr:RobustUniformReleaseProb}
  For all $L = k-1, \ldots, R$, $K\cdot r(L) \geq L/(L-\sigma R)$ for a suitable constant $\sigma \geq 0$, excluding the contribution of $\tau(k/R)$.
\end{proposition}

The parameter $L$ represents the number of unsolved input symbols, $r(L)$ is the overall release probability that an output symbol is released when $L$ input symbols remain unsolved and $K$ is the number of output symbols. The derivations of the expressions of these parameters can be found in~\cite{Luby:02:inproceedings}. Here we apply the same idea to show the release probability of proposed scheme in phase one. From the expression of the number of output symbols $K$ in LT codes, we can derive the number of output symbols, $K_{mclt_1} = (\gamma/\beta)K$, needed in phase one in our 2-configuration LT codes and it will also have the robust release probability.\par

\begin{table*}[t!]
\caption{The table lists the detailed settings of the three experiments. The overhead is an average over $2 \times 10^6$ independent runs.}
\begin{center}
\begin{tabular}{|c||c|c|c|}
\hline
Label & Robust & Starter & Starter+Closer \\ \hline
Method & LT codes & LT codes & $ \!\! $ 2-configuration LT codes $ \!\! $ \\ \hline
Distribution & robust soliton & starter & starter+closer \\ \hline
Parameters & $ \!\! c=0.1$, $\delta=0.1 \!\! $ & $ \!\! c=0.1$, $\delta=0.1$ \!\! & $c=0.1$, $\delta=0.1$ \\ \hline
Switch & - & - & $R=c \ln(k/\delta)\sqrt{k}$ \\ \hline \hline
Overhead & 0.2416 & 0.1288 & 0.0932 \\ \hline
\end{tabular}
\label{t:experiments}
\end{center}
\end{table*}

\begin{proposition}
  \label{pr:RobustUniformReleaseProb_MCLT}
  For all $L = k-1, \ldots, R$, $K_{mclt_1}\cdot r_{mclt_1}(L) \geq L/(L-\sigma R)$ for a suitable constant $\sigma \geq 0$.
\end{proposition}

The proof of this proposition is based on the proof of Proposition~\ref{pr:RobustUniformReleaseProb}. Since it only used the contributions of $\tau(2), \ldots, \tau(k/R-1)$ and the ideal soliton distribution $\rho(\cdot)$, we can reduce $K_{mclt_1}\cdot r_{mclt_1}(L)$ to $K\cdot r(L)$.
\begin{align*}
  &K_{mclt_1}\cdot r_{mclt_1}(L) \\
  =& \frac{\gamma}{\beta}K\cdot \sum_i r_{mclt_1}(i, L) \\
  =& \frac{\gamma}{\beta}K\cdot\sum_i\left(\lambda(i)\cdot q(i, L)\right) \\
  =& \frac{\gamma}{\beta}K\cdot\sum_i^{k/R-1}\left(\frac{\rho(i)+\tau(i)}{\gamma}\cdot q(i, L)\right) 
  + \frac{\gamma}{\beta}K\cdot\sum_{k/R}^k\left(\frac{\rho(i)}{\gamma}\cdot q(i, L)\right) \\
  =& K\cdot\sum_i\left(\frac{\rho(i)+\tau(i)}{\beta}\cdot q(i, L)\right) \; , \; \mbox{excluding the contribution of $\tau(k/R)$} \\
  =& K\cdot r(L)
\end{align*}
According to Proposition~\ref{pr:RobustUniformReleaseProb}, it shows that the proposed 2-configuration LT codes also has the same robust release probability. We can also derive the expected number of output symbols of degree one in phase one:
\begin{align*}
  K_{mclt_1}\cdot \lambda(1) &= \frac{\gamma}{\beta}K\cdot\left(\frac{\rho(1)+\tau(1)}{\gamma}\right) \\
			    &= K\cdot\left(\frac{\rho(1)+\tau(1)}{\beta}\right) \\
                &\approx R \; .
\end{align*}

Therefore, the phase one of the 2-configuration LT codes would have the same expected number of output symbols of degree one and the same robust uniform release probability. It shows that the decoding process would be successful until $R$ input symbols remain unprocessed with a probability at least $1-\delta/3$, as the original LT codes.\par

When the number of unsolved input symbols is down to $R$, the receiver can change to receive output symbols of degree $k/R$ generated by the second configuration. According to the degree domination relation described by Equation~(\ref{eq:DegreeDominationRelation}), when the number of unsolved input symbols is less than $R$, the output symbols of degrees lower than $k/R$ are dominated by those of degree $k/R$. Therefore, the closing distribution can help receivers to recover remaining input symbols more efficiently. Applying the same concept as the proof of the robust release at end probability by Luby, the closing distribution can cover the second decoding phase with high probability when a sufficient amount of output symbols is received. In summary, since we design the 2-configuration LT codes based on the idea of robust soliton distribution, the proof of the bound of decoding failure rate of robust soliton distribution also works for the proposed design when approximately the same number of output symbols are received from the starting and closing distribution with a reasonable proportion. Accounting for the scheme with multiple configurations provides the receivers the ability to collect more helpful output symbols at different decoding phases which acts like rearranging the output symbols into a better receiving order corresponding to the decoding process, the performance of multiple configuration LT codes is expected better than pure LT codes in most cases. We will present the experimental results to confirm the improvement in the next section.\par

\section{Experiments}
\label{sec:Experiments}

In section~\ref{subsec:ArbitraryMCLT}, an approach to create a pair of degree distributions was introduced to cooperate with MC-LT codes. The design of the pair of degree distributions is inspired by robust soliton distribution, and MC-LT codes with such a design, in theory, perform as good as pure LT codes for the extreme case. In order to confirm the improvement in practice of the MC-LT codes, the simulation experiments are conducted to compare the performance of MC-LT codes and pure LT codes in this section. There are three experiments conducted. The detailed settings are given in Table~\ref{t:experiments}. The first and second experiments are pure LT codes with different degree distributions. The last one is MC-LT codes with the proposed degree distribution pair. \textit{Robust}, \textit{Starter}, and \textit{Starter+Closer} are used respectively to indicate the three subjects in this study. Aiming at broadcasting scenarios for real-time multimedia streaming, the code length in the experiments is set as $k=1000$. Distribution parameters, $c=0.1$ and $\delta=0.1$, are identical in all experiments. $R$ is the switch criterion required only by MC-LT codes. The results of reception overhead for a full recovery are recorded at the bottom of Table~\ref{t:experiments}. The numbers of overhead are all averaged of $2 \times 10^6$ independent runs of simulations. It is clear that \textit{Starter+Closer} and \textit{Starter} outperform \textit{Robust} for this performance indicator.\par

\begin{figure}[t!]
\centerline{\includegraphics[width=90mm]{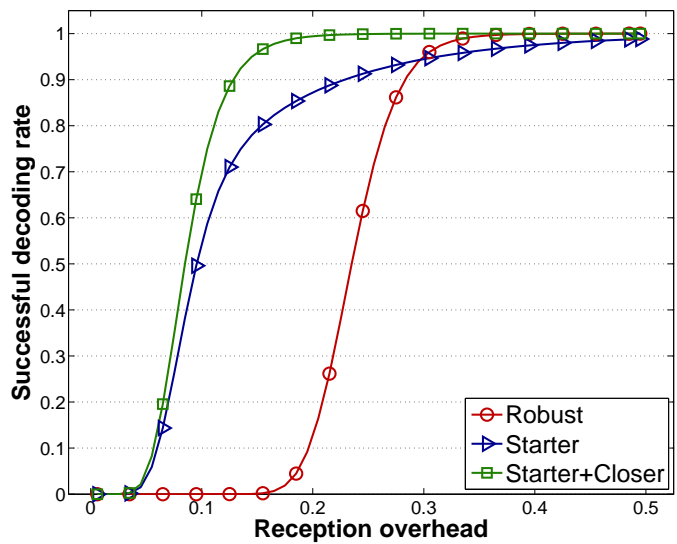}}
\caption{The figure shows the successful recovery rate for the three experiments.}
\label{fig:success_rate}
\end{figure}

The overall information is illustrated in Figure~\ref{fig:success_rate}. There is no doubt that \textit{Starter+Closer} is the best among the three subjects. LT codes with starting distribution seem to work better than LT codes with robust soliton distribution. The improvement of reception overhead is simply obtained by replacing the adopted degree distribution. As shown in Figure~\ref{fig:success_rate}, \textit{Starter} does not dominate \textit{Robust}, although the average reception overhead of \textit{Starter} is much lower. The reason can be found in Figure~\ref{fig:MC_dis}, which shows that more probability is distributed to lower degrees after the peak $\tau(k/R)$ is removed from the robust soliton distribution. Distribution with a lower average degree is advantageous of employing belief propagation for decoding. \textit{Starter} and \textit{Starter+Closer} consequently have a better chance to complete the decoding process with less redundancy. However, potential fractions of isolated input symbols is the critical weakness of LT codes with a low average degree distribution. Isolated input symbols are never chosen to take part in the encoding process such that it is impossible to recover them at the receiver side. It can be found that \textit{Starter} encounters a difficulty in achieving $100\%$ successful decoding rate even if the overhead is $0.5$.\par

\begin{figure}[t!]
\centerline{\includegraphics[width=90mm]{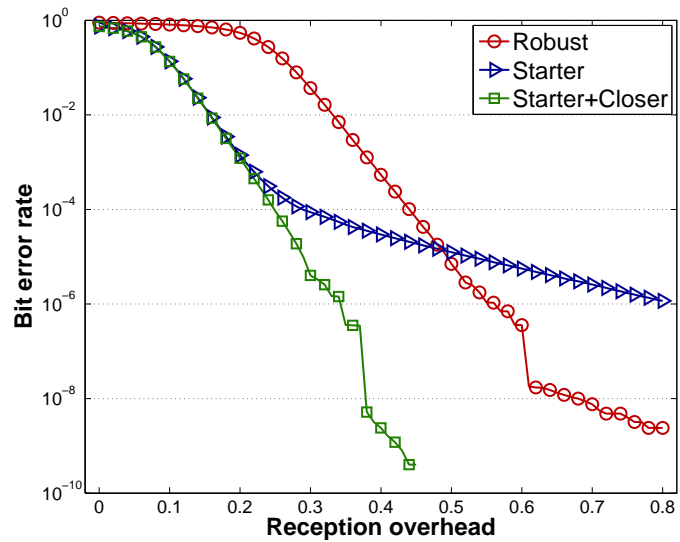}}
\caption{The figure displays the bit error rate for the three experiments.}
\label{fig:error_rate}
\end{figure}

Figure~\ref{fig:error_rate} illustrates the experimental results with a different performance indicator. Bit error rate of each experiment subject is plotted in the log scale. The results not only confirm that \textit{Starter+Closer} is the best but also show the reason why MC-LT codes work well. Starting distribution effectively helps the decoding process in early phases and consequently the bit error rates of \textit{Starter} and \textit{Starter+Closer} drop quickly. The two curves are almost identical until the reception overhead reaches $0.2$. The moment is at the point that the remaining unsolved input symbols are fewer than $R$, and \textit{Starter+Closer} therefore switches to receive output symbols from the second configuration. Closing distribution then provides output symbols of a high degree to continuously reduce the error rate. In contrast, \textit{Starter} evidently failed without the help of closing distribution. The simulation results show the significant improvement of MC-LT codes and also indicate that the proposed approach is practical for constructing distributions for MC-LT codes.\par

\section{Conclusions}
\label{sec:Conclusions}


In this work, we introduced the scheme of the multiple configurations LT codes to improve the performance of LT codes. The proposed scheme aims at the broadcasting scenarios without feedback channels and trades in the cost at the sender side for performance. The sender of MC-LT codes generates multiple sets of output symbols with different configurations. The receivers gain the benefit by collecting more useful set of output symbols depending on the local decoding phases. The MC-LT codes preserve the advantages of LT codes, such as the universal property and being suitable for broadcasting. We derived the degree domination relation, the foundation of MC-LT codes, and presented examples to demonstrate the improvement over the optimal pure LT codes. A practical design of MC-LT codes for arbitrary code length $k$ is also given in this paper. In theory for the extreme case, MC-LT codes with the proposed distributions have the same performance bound as pure LT codes with robust soliton distribution. The simulation experiments demonstrate the significant performance improvement of MC-LT codes in practice.\par

This paper presents a proof-of-principle study and augments the research scope of LT codes by letting LT codes users make decisions on the following three new factors: 1) How many configurations should be used? 2) When should a receiver switch from one configuration to another? 3) What degree distribution or encoding setting should be used for each configuration? This study has shown that with appropriate settings, MC-LT codes can indeed outperform the optimal pure LT codes, while the optimal MC-LT codes have yet to be discovered. All these items should be further investigated. We will continue to conduct research along this line and also take the cost at sender side into consideration to strike a balance between the number of configurations and the overall transmission performance.\par




\bibliographystyle{IEEEtran}
\bibliography{IEEEabrv,MCLT}

\begin{thebibliography}{1}
\providecommand{\url}[1]{#1}
\csname url@samestyle\endcsname
\providecommand{\newblock}{\relax}
\providecommand{\bibinfo}[2]{#2}
\providecommand{\BIBentrySTDinterwordspacing}{\spaceskip=0pt\relax}
\providecommand{\BIBentryALTinterwordstretchfactor}{4}
\providecommand{\BIBentryALTinterwordspacing}{\spaceskip=\fontdimen2\font plus
\BIBentryALTinterwordstretchfactor\fontdimen3\font minus
  \fontdimen4\font\relax}
\providecommand{\BIBforeignlanguage}[2]{{%
\expandafter\ifx\csname l@#1\endcsname\relax
\typeout{** WARNING: IEEEtran.bst: No hyphenation pattern has been}%
\typeout{** loaded for the language `#1'. Using the pattern for}%
\typeout{** the default language instead.}%
\else
\language=\csname l@#1\endcsname
\fi
#2}}
\providecommand{\BIBdecl}{\relax}
\BIBdecl

\bibitem{Luby:02:inproceedings}
M.~Luby, ``{LT} codes,'' in \emph{Proc. 43rd annual IEEE Symposium on
  Foundations of Computer Science}, 2002, pp. 271--280.

\bibitem{Byers:98:inproceedings}
J.~W. Byers, M.~Luby, M.~Mitzenmacher, and A.~Rege, ``A digital fountain
  approach to reliable distribution of bulk data,'' in \emph{Proc. of the ACM
  SIGCOMM '98 Conference on Applications, Technologies, Architectures, and
  Protocols for Computer Communication}, 1998, pp. 56--67.

\bibitem{He:10:inproceedings}
N.~He, J.~Cao, Z.~Li, and Y.~Ren, ``{JVEC}: Joint video adaptation and erasure
  code for wireless video streaming broadcast,'' in \emph{Proc. of 2010 IEEE
  International Conference on Communications}, 2010, pp. 1--5.

\bibitem{Ji:11:inproceedings}
W.~Ji and Z.~Li, ``Heterogeneous {QoS} video broadcasting with optimal joint
  layered video and digital fountain coding,'' in \emph{Proc. of 2011 IEEE
  International Conference on Communications}, 2011, pp. 1--6.

\bibitem{Bonello:09:article}
N.~Bonello, R.~Zhang, S.~Chen, and L.~Hanzo, ``Reconfigurable rateless codes,''
  \emph{IEEE Transactions on Wireless Communications}, vol.~8, no.~11, pp.
  5592--5600, 2009.

\bibitem{Hyytia:07:inproceedings}
E.~Hyyti{\"{a}}, T.~Tirronen, and J.~Virtamo, ``Optimal degree distribution for
  {LT} codes with small message length,'' in \emph{Proc. of the 26th IEEE
  International Conference on Computer Communications ({INFOCOM} 2007)}, 2007,
  pp. 2576--2580.

\end{thebibliography}

\end{document}